\documentclass[prl,twocolumn,showpacs,superscriptaddress]{revtex4-1}
\usepackage{graphicx,amssymb,amsmath,psfrag,color}
\begin{document}
\definecolor{darkgreen}{rgb}{0,0.5,0}
\newcommand{\be}{\begin{equation}}
\newcommand{\ee}{\end{equation}}

\title{Loschmidt echo and the many-body orthogonality catastrophe in a qubit-coupled Luttinger liquid}

\author{Bal\'azs D\'ora}
\email{dora@eik.bme.hu}
\affiliation{BME-MTA Exotic  Quantum  Phases Research Group, Budapest University of Technology and
  Economics, Budapest, Hungary}
\affiliation{Department of Physics, Budapest University of Technology and Economics, Budapest, Hungary}
\author{Frank Pollmann}
\affiliation{Max-Planck-Institut f\"ur Physik komplexer Systeme, Dresden, Germany}
\author{J\'ozsef Fort\'agh}
\affiliation{CQ Center for Collective Quantum Phenomena and their Applications, Eberhard-Karls-Universit\"at T\"ubingen, T\"ubingen, Germany}
\author{Gergely Zar\'and}
\affiliation{BME-MTA Exotic  Quantum  Phases Research Group, Budapest University of Technology and
  Economics, Budapest, Hungary}

\date{\today}

\begin{abstract}
We investigate the many-body generalization of the orthogonality catastrophe by studying the 
generalized Loschmidt echo of Luttinger liquids (LLs) after  a global change of interaction. 
It decays  exponentially with system size and exhibits universal behaviour: 
the steady state exponent after quenching back and forth $n$-times between 2 LLs (bang-bang protocol)
is $2n$-times bigger than that of the adiabatic overlap, and depends only on the initial and final
 LL parameters. 
These are corroborated numerically by  matrix-product state based methods of the 
 XXZ Heisenberg model.
An experimental setup consisting of a hybrid system containing cold atoms 
and a flux qubit coupled to a Feshbach resonance is proposed to measure the Loschmidt echo using rf spectroscopy or Ramsey interferometry. 
\end{abstract}


\maketitle

{\em Introduction.} 
The long coherence times and the possibility to control parameters very accurately 
in optical lattices allow to simulate exciting non-equilibrium effects in quantum many body systems. Particular attention has been devoted to the evolution of quantum
stated after quenches\cite{polkovnikovrmp} in which a parameter in the Hamiltonian is changed either suddenly or gradually. By measuring the Loschmidt echo (LE), it 
is possible to get insight into the dynamical properties of the quantum-many body state.

The LE is defined as the overlap of two wave functions, $|\Psi_0(t)\rangle$ and $|\Psi(t)\rangle$, evolved from the same
initial state,  but with different Hamiltonians, $H_0$ and $H$,
\begin{equation} 
\mathcal{L}(t)\equiv \left|\langle\Psi_0(t)|\Psi(t)\rangle\right|^2.
\end{equation}
It measures the "distance" between two quantum states and serves to
 quantify irreversibility and chaos in quantum mechanics\cite{peres,physrep,goussev}.
Furthermore, it can be used to diagnose quantum phase transitions\cite{pollmann},
and is also  an important quantity in various fields of physics, ranging from 
 nuclear magnetic resonance  to quantum computation and information theory.
For the latter, the LE is of utmost importance since it measures how small changes during a time evolution cause decoherence and are detrimental for
quantum information processing and storage. 

While the theoretical concept behind the LE is very clear, 
experimental protocols to measure the overlap of two wave functions, are challenging and rare. 
Experimentally, the LE and the closely related fidelity has only been observed in systems with limited degrees of freedom~\cite{zhang2}, 
by  coupling the system of interest to  a single qubit, whose states probe the wavefunction of the system at different values of some external 
parameters. Also, while the LE for local perturbations --  the 
X-ray edge singularity problem -- is well understood, its behavior  in generic many-body systems is poorly described~\cite{venuti2}. 
Studies so far mostly focused on local perturbations and simple refermionizable spin chain models\cite{goold,knap,rossini}.

\begin{figure}[b!]
\includegraphics[width=6cm]{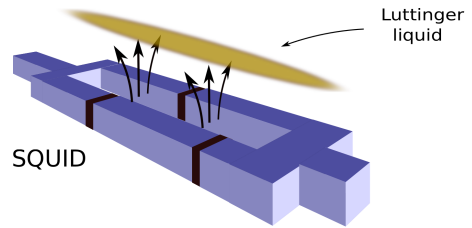}
\caption{(Color online) Schematics of the experimental setup. The black segments on the SQUID denote the Josephson junctions, the arrows stand for the total magnetic field.
By changing the eigenstate of the flux qubit,
the total flux hence the total magnetic field changes, controlling the Feshbach resonance in the cold atomic LL.}
\label{cartoon}
\end{figure}

Here we study the interaction driven LE of a genuine  interacting one-dimensional system, a Luttinger liquid (LL). 
One-dimensional strongly correlated systems often form such Luttinger liquids (LL), 
and can  also be  created from cold atoms~\cite{cazalillarmp}.
Recently, the LL wavefunction has been  used to evaluate the overlap of distinct LL ground states \cite{fjaerestad,yang}, and
optimal protocols producing maximal overlaps with reference states have  also been investigated~\cite{rahmani}.
In this work,  we evaluate the LE within the Luttinger liquid (LL) description for a time dependent Hamiltonian,  
and verify the predictions of the LL model numerically on the XXZ Heisenberg chain using matrix-product state (MPS) based methods. 
Furthermore, we propose an experimental setup to measure the LE  of a Luttinger liquid, where  a flux qubit coupled 
to a Feshbach resonance is used to control the interaction in one-dimensional cold atom gas (see Fig.~\ref{cartoon}).

{\em Model.} A LL can be described in terms of bosonic  sound-like collective excitations, 
regardless to the  statistics of the original system. The LL 
Hamiltonian is given by\cite{giamarchi}
\begin{equation}
H_0=\sum_{q\neq 0}\left(  \omega_q a_q^+ a_q
+\frac{g_i(q)}{2}[a_qa_{-q}+a_q^+a_{-q}^+]\right) \;,
\label{eq:H_0}
\end{equation}
where $g_i(q)$ is the initial interaction, and $\omega_q$ the energy of bosonic excitations. In the following, we shall assume that $H=H(t)$ has the same form as 
Eq. \eqref{eq:H_0}, but with a time dependent coupling, 
\be
g_q\to g_q(t) = g_i({q}) + \Delta g_q(t),
\ee
where $\Delta g_q(t)=[g_i({q})-g_f({q})] Q(t)$, and $g_f({q})$ is the final interaction strength.  For a conventional LE $Q(t)\equiv 1$ for 
$t>0$, but here we allow any time dependence, and only assume that $Q(0)=0$ and $Q(t)=1$ after some "transition time"
$\tau$ ~\footnote{This characterizes intentional (given $Q(t)$ protocol) or unavoidable (finite duration of pulses) timescales.}.

The Hamiltonian $H(t)$ is quadratic, and can be diagonalized at any instance. Its initial and final quasiparticle  spectra
are simply given by 
$\omega_{i/f}({q})=(\omega^2_q- g_{i/f}^2({q}))^{1/2}$, 
and the strength of interaction 
in these states is conveniently characterized by the dimensionless LL parameters, 
$K_{i/f}=\sqrt{\left[{\omega_q-g_{i/f}({q}) }\right]/\left[{\omega_q+g_{i/f}({q})}\right]}$.

To compute the LE, it is convenient to first diagonalize  Eq.~\eqref{eq:H_0} 
 by a standard, time independent Bogoliubov transformation,
yielding $H_0=\sum_{q\neq 0}  \omega_i({q})\; b_q^+ b_q$ up to a constant shift of energy.
In this basis,  $H(t)$ reads
\begin{gather}
H= \sum_{q\neq 0}\omega(q,t)b_q^+ b_q+\frac{ g(q,t)}{2}[b_qb_{-q}+b_q^+b_{-q}^+]+\dots
\label{hamilton}
\end{gather}
with $\omega(q,t)=\omega_i({q})-\Delta g_q(t)\frac{g_i({q})}{\omega_i({q})}$ and 
$ g(q,t)={\Delta g_q(t)}\frac{\omega_q}{\omega_i({q})}$, and the dots stand for  an unimportant  time dependent energy shift. 
We emphasize that our setting is very general ~\footnote{For simplicity, we have neglected velocity renormalization by the interactions (e.g. $g_4$ type process\cite{giamarchi}), 
which would only make the algebra more involved, but does not affect our results.}, and equally applies for a  spinless fermion, boson or spin system\cite{giamarchi}. 

%
%

{\em Analytical results.} In order to calculate the LE\cite{silva}, we need to determine the wave function of our quenched LL. This can
be achieved by realizing that Eq. \eqref{hamilton} couples only pairs of states with $q$ and $-q$. Consequently, 
the total
time-evolution operator is  the product of the separate evolution operators $U_q(t)$  for all $q$'s.
Finding $U_q(t)$ for a given pair of modes can be accomplished  by realizing that the operators appearing in the Hamiltonian, 
$K_0(q)=(b^+_qb_{q}+b_{-q}b^+_{-q})/{2}$, $K_+(q)=b^+_qb^+_{-q}$ and $K_-(q)=b_qb_{-q}$
 are the generators of  SU(1,1) Lie algebra.
Generalizing the results of squeeze operators to the present case, 
and exploiting a faithful matrix representation of the SU(1,1) generators ~\cite{gilmore}, the time evolution operator $U_q(t)$ can finally be expressed as~\cite{EPAPS} 
\begin{gather}
U_q(t)=\exp[C_+(q,t) K_+(q)]\exp[C_0(q,t) K_0(q)]\times\nonumber\\
\times\exp[C_-(q,t)K_-(q)+i\varphi_q(t)].
\label{timeevolop}
\end{gather}
Here $\varphi_q(t)$ is an unimportant phase, and  the coefficients $C_0(q,t)$ and $C_{\pm}(q,t)$ can be expressed in terms of the 
time dependent Bogoliubov coefficients, $u_q(t)$ and $v_q(t)$, 
\begin{subequations}
\begin{gather}
C_+(q,t)={v_q^*(t)}/{u_q^*(t)},\hspace*{2mm} C_-(q,t)=-{v_q(t)}/{u_q^*(t)},\\
C_0(q,t)=-2\ln(u_q^*(t)), 
\end{gather}
\end{subequations}
with the coefficients $u_q(t)$ and $v_q(t)$ satisfying \cite{EPAPS}
\begin{gather}
i\partial_t\left[\begin{array}{c}
u_q(t)\\
v_q(t)\end{array}\right]=\left[\begin{array}{cc}
\omega(q,t) & g(q,t)\\
-g(q,t) & -\omega(q,t)
\end{array}\right]
\left[\begin{array}{c}
u_q(t)\\
v_q(t)\end{array}\right]
\label{beq}
\end{gather}
with $[u_q(0),v_q(0)]=[1,0]$.
We can now express the time evolution of any initial state $|\Psi_0\rangle $ as  $|\Psi(t)\rangle=U(t)|\Psi_0\rangle$, 
with $U(t)=\prod_{q>0} U_q(t)$.

In the following, for simplicity, we take $|\Psi_0\rangle $ as the ground state wavefunction of $H_0$, which is the vacuum for the $b$ bosons.
Then  the time evolved, normalized wave function simplifies to
\begin{gather}
|\Psi(t)\rangle=e^{-i\Phi(t)+\sum_{q>0}\bigl[ \frac{v_q^*(t)}{u_q^*(t)} b^+_qb^+_{-q} 
-\ln[u_q^*(t)]\bigr]\vphantom{\sum_i}}|\Psi_0\rangle,
\label{wf}
\end{gather}
with $\Phi(t)$ an overall phase factor. This 
generalizes the equilibrium results\cite{fjaerestad,yang} to the non-equilibrium situation, encoded via the time 
dependent Bogoliubov coefficients. 
Then the time evolution under the action of   $H_0$ is trivial,  since $|\Psi_0\rangle$
only picks up a phase, and the LE takes a particularly simple form
\begin{gather}
\mathcal{L}(t)=\exp\left(-\sum_{q>0}\ln\left(|u_q(t)|^2\right)\right).
\label{le}
\end{gather}
This result, in combination with Eq. \eqref{beq} determines the complete time dependence of the generalized LE in a LL. 
It holds true for any non-equilibrium evolution, and expresses the LE  solely in terms of   the number of excited quasiparticles 
in the final state. 
We regularize the  $q$ sums in \eqref{le}  by an $\exp(-\alpha |q|)$ factor, with $1/\alpha$ an ultraviolet cutoff\cite{giamarchi}.

We can now use the function $Q(t)$ to calculate the LE. Changing $Q(t)$ adiabatically, 
one obtains from Eq. \eqref{beq} $|u^{\rm ad}_q(t)|^2=1/2+ \left({K_i}/{K_f}+{K_f}/{K_i}\right)/4$
for times $t>\tau$. The LE  is in this case just the overlap of the ground states 
of the initial and final Hamiltonian and, in agreement with Refs. \cite{fjaerestad,yang}, reads as
\begin{gather}
\mathcal L_{ad}=\left(\frac 12+\frac 14\left(\frac{K_i}{K_f}+\frac{K_f}{K_i}\right) \right)^{-L/2\pi\alpha},
\label{lead}
\end{gather}
with $L$ being the system size. This remains valid in the steady state for near-adiabatic quenches, $\tau\gg \alpha/v$ with $v$ being the
sound velocity in the final state.

\begin{figure}[h!]


\includegraphics[width=8cm]{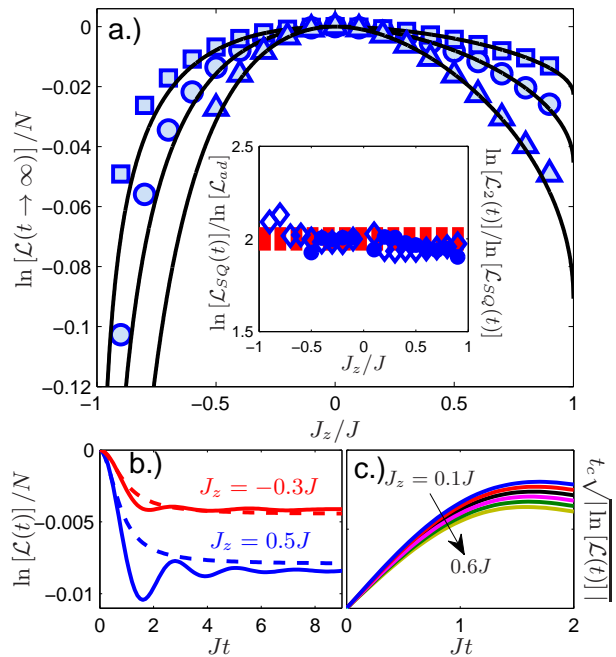}

\caption{(Color online) a.) The exponent of the generalized Loschmidt echo is shown for SQ $n=1$ (circle), double quench $n=2$ (triangle)
 and adiabatic time evolution (square) for the XXZ Heisenberg model
in the steady state,
starting from the XX point and ending up at a finite $J_z$, obtained numerically from MPS based methods (when convergence was reached).
The solid lines are the analytical results from
Eqs. \eqref{lead} and
\eqref{leSQ}. 
The inset shows the ratio of the SQ and adiabatic exponents (diamond) and the $n=2$ and 1 exponents (circle)
from numerics, which agrees well with the theoretically expected value of 2.
b.)  Typical  numerical results (solid lines) of the LE of the XXZ model are shown for a SQ from the XX point to $J_z=0.5J$ and -0.3$J$,
the dashed curve is the analytical expression using  Eqs. \eqref{le} and \eqref{uSQ}.
c.) The scaling of the numerical data  expected from Eq.
\eqref{transient} for short times for a SQ is visualized from $J_z=0$ to $0.1J$ to $0.6J$ with $0.1J$ steps from top to bottom in arbitrary units.
}
\label{lexxz}
\end{figure}

For $Q(t>0)=1$, i.e., a conventional "sudden quench" (SQ)~\cite{cazalillaprl} LE,  however, Eq. \eqref{beq} yields
\begin{gather}
|u^{\rm}_q(t)|^2=1+\frac 14 \sin^2(\omega_f({q})t)\left(\frac{K_f}{K_i}-\frac{K_i}{K_f}\right)^2,
\label{uSQ}
\end{gather}
with $\omega_f(q)=v|q|$ denoting the excitation energy after the quench. 
By plugging this back to Eq. \eqref{le}, we find for the short time limit ($t\ll \alpha/v$)  
\begin{gather}
\mathcal L_{SQ}(t)\sim   \exp\left(-c\; L\; (t/t_c)^2/\alpha\right)
\label{transient}
\end{gather}
where $c$ is a non-universal constant of order unity. The characteristic decay time of this expression is 
$t_c\equiv 4\alpha/v|K_f/K_i-K_i/K_f|$, and $1/t_c^2 \sim$ can be identified 
as the  variance of energy per particle after the quench per particle.   For intermediate times $t\sim t_c$, the LE displays a non-universal transient signal 
(see Fig.~\ref{lexxz}b), 
however, for very long times, $t\gg \alpha/v$,  the LE for  SQ  becomes time independent and universal:
\begin{gather}
\mathcal L _{SQ}(t\gg\alpha/v)=\left(\frac 12+\frac 14\left(\frac{K_i}{K_f}+\frac{K_f}{K_i}\right) \right)^{-L/\pi\alpha},
\label{leSQ}
\end{gather}
which holds also true for fast quenches, $\tau\ll\alpha/v$.
Excitations are only produced at $t=0$, which interfere with each other for a short amount of time, causing Eq. \eqref{transient},
but after this phase coherence is lost, excitations propagate independently and only their total number determines the overlap.

By Eq.~\eqref{lead} we just established that for $t\to \infty$
\begin{gather}
 \mathcal  L_{SQ} = \mathcal L_{ad}^2.
\label{leuniv} 
\end{gather}
This can be understood in terms of simple physical arguments and must hold in general, too\cite{rossini}:
the adiabatic LE measures just the square of the overlap $\langle G|G_0\rangle$ of the ground state of the initial 
and final Hamiltonians, which is  the probability weight of the ground state of $H$ in $|G_0\rangle=|\Phi_0\rangle$. 
For a regular LE, $ \langle\Phi(t)|\Phi_0(t) \rangle = \langle\Phi_0|  e^{i t H} e^{-i t H_0}|\Phi_0 \rangle$. 
Inserting here the complete set of eigenstates of   $H$ between the two propagators, the excited states
amount in oscillating terms,
interfering destructively, and only the ground state contribution remains, yielding  asymptotically in the thermodynamic limit
$ |\langle\Phi(t)|\Phi_0(t) \rangle| =  | \langle G|G_0\rangle|^2$, 
and thus $ \mathcal L_{SQ}(t)=|\langle\Phi(t)|\Phi_0(t) \rangle|^2\xrightarrow{t\rightarrow\infty}| \langle G|G_0\rangle|^4 = \mathcal L_{ad}^2  $ .
Using the analogy to work statistics\cite{silva}, this is the square of the probability to stay in the ground state after the quench.

%
%

The exponent of the LE is further enhanced\cite{EPAPS} by repeating the
$Q(t>0)$:  SQ to 1, holding time $t$, SQ to 0, holding time $t$
sequence  $n$ times (bang-bang protocol).
The generalized LE is the $n$th power of the SQ overlap in Eq. \eqref{leSQ} as
$\mathcal L_{n}(2nt)=\mathcal L_{SQ}^{n}(t)=\mathcal L_{ad}^{2n}$ in
the post-quench steady state, therefore
the exponent is enhanced by a factor of $n$.

{\em Numerics.} We tested the analytical predictions numerically on the 
one-dimensional XXZ Heisenberg model\cite{giamarchi}, covering $1/2<K<\infty$, for
 adiabatic ramps and  SQ from $J_z=0$ to a finite $J_z$. We  performed the simulations using a
combination of MPS \cite{Fannes-1992} based infinite density matrix renormalization group
\cite{White-1992,Kjaell-2012,McCulloch-2008} to find the ground state variationally, 
and the infinite time evolving block decimation \cite{Vidal-2007}
algorithms to simulate the quench.
The LE is  calculated by finding the dominant eigenvalue of a "generalized" transfer 
matrix which we obtain by contracting the tensors representing the two states. 
 
The factor $L/\pi\alpha$ in the exponent in Eqs.  \eqref{lead} and \eqref{leSQ} 
contains the unknown short distance cutoff. However, its value can be fixed by calculating the fidelity 
susceptibility, $\chi_f$, around the non-interacting XX point of the Heisenberg model, in which case 
$L/2\pi\alpha\approx N\chi_f\pi^2$, where $N$ is the number of lattice 
sites and $\chi_f\approx 0.0195$\cite{sirker}. Using then the Bethe Ansatz result~\cite{giamarchi},
$K=\pi/2[\pi-\arccos(J_z/J)]$, we find an excellent  agreement  with the numerical data 
with \emph{no} fitting parameter (see Fig. \ref{lexxz}a).  
This excellent 
agreement is somewhat  surprising since, as expected,  the non-universal transient signals 
clearly differ in the LL approach and the numerics, and also, because  
the LL description completely neglects (asymptotically irrelevant) back  scattering processes,  
contained in the the lattice calculations. 

Slight deviations  show up only at the end points of the critical region of the XXZ Heisenberg model in Fig. \ref{lexxz}.
Close to the $J_z=J$ point, the previously mentioned back scattering term, driving the Kosterlitz-Thouless phase transition
causes a slight disagreement. Upon approaching the ferromagnetic critical point at $J_z=-J$, on the other hand, 
the validity of bosonization shrinks to very small energies, and the high energy modes,
absent in the present considerations, also influence the overlap.

{\em Detection:}
We propose to measure  the LE of the LL  in a cold atomic setting,  where a flux qubit\cite{friedman,vanderWal} is
used   to control the interaction between a the atoms. 
{  Though here we focus on a one dimensional LL, the proposed setup also opens up 
the possibility to study the LE  in higher dimensional interacting systems.}
The flux qubit consists  of a Josephson junction circuit, 
and is governed by the Hamiltonian $H_{\rm qubit}=\epsilon\sigma_z+\Delta \sigma_x$, with $\Delta$ the tunneling between the 
two eigenstates of $\sigma_z$, 
$|\circlearrowleft\rangle$ and $|\circlearrowright\rangle$, carrying oppositely circulating persistent currents, $\pm I$, 
and $\epsilon$ the energy splitting. 
In addition to the external flux, $\Phi_{\rm ext}$, the states $|\circlearrowleft\rangle$ and $|\circlearrowright\rangle$
generate an additional flux, $\Phi_f$. Ideally, tunneling between them is suppressed. 
%
The one-dimensional quantum gas is prepared in a two-dimensional optical lattice potential\cite{Greiner2001} at the chip surface, hosting the flux qubit.  A few
microns thick, high reflectivity ($>$99.9\%) dielectric coating of the chip { (not shown in Fig.~\ref{cartoon})}  can shield the superconducting device from the laser power (mW) incident during the measurement time (ms).
The quantum gas is positioned at the flux qubit (see Fig.~\ref{cartoon}) by positioning the laser beams (optical tweezers), such that the total magnetic field 
of the state $|\circlearrowright\rangle$  of flux $\Phi_{\rm ext}+\Phi_f$
be at a Feshbach resonance, while the field of the state  $|\circlearrowleft\rangle$ of flux $\Phi_{ext}-\Phi_f$, be further away  from the resonance.
We estimated the qubit switching-induced magnetic field difference for   an elongated rectangular flux qubit, with parallel sides
 comparable to the length of a typical cold atomic tube  ($\sim 10~\mu$m). Assuming 
 a persistent current of $I=2\mu$A and a separation of $d=2R = 2\mu$m between the two lines of the qubit,
 we obtain a field difference $\delta B_f\sim16$~mG. Although relatively small, this field is comparable 
 to the width $\Delta B=15~$mG of some narrow Feshbach resonances used to realize 
 a LL in   $^{87}$Rb systems~\cite{widera}. 
 

In this setup, one could  use rf spectroscopy and
measure the absorption spectrum of the
qubit~\cite{Pothier} in the presence and in the absence of the trapped gas.
Similar to the X-ray edge singularity problem, this absorption signal is just proportional
to the Fourier transform of the LE.
{ 
In this case, the qubit can be positioned far away from the degeneracy point, at $|\epsilon| \gg \Delta$,
and the rf/mw spectroscopy 
can be referenced to high stability quartz oscillators ($\delta f/ f  \sim 10^{-13}$) to resolve the fine structure at the absorption edge. 
}

Alternatively, the LE can  be measured   using Ramsey interferometry\cite{knap,goold}:
initializing the qubit in the $|\circlearrowleft\rangle$ state with weak interactions to the cold atoms, 
yields a  wavefunction  $|\circlearrowleft\rangle\otimes |\Psi_0\rangle$ at $t=0$. By applying a $\pi/2$ 
rf pulse~\footnote{One can create such transitions even for $\Delta\approx0$ 
using off resonance multiphoton processes.}, a superposition
of the two qubit states is produced $(|\circlearrowleft\rangle+|\circlearrowright\rangle)/\sqrt 2\otimes |\Psi_0\rangle$,
yielding distinct, qubit state dependent time evolution for $|\Psi_0\rangle$. 
After time $t$, a second $\pi/2$ pulse and the measurement of the qubit current $\langle\hat I \rangle\sim\langle\sigma_z\rangle $ 
is performed, giving a signal  proportional to ${\cal L}(t)$.  
The timescale separating Eqs. \eqref{leSQ} and \eqref{transient} is estimated as 
$\alpha/v\sim 40~\mu$s from a typical Fermi temperature of 1~$\mu$K
or trapping frequency of $25$~kHz\cite{widera}, and $t_c$ can be made even shorter by executing large quenches.
Given the rapid progress of technology of flux qubits\cite{norirmp}, these are already accessible with current 
coherence times and will be even more so in the future.


{\em Note added.} After accomplishing this work, we became aware of related proposals to measure the work statistics in hybrid architectures\cite{mazzola,dorner}.

\begin{acknowledgments}

We acknowledge useful discussions with   M. Haque.
This research has been  supported by the Hungarian Scientific  Research Funds Nos. K101244, K105149, CNK80991,
T\'{A}MOP-4.2.1/B-09/1/KMR-2010-0002  and by the ERC Grant Nr. ERC-259374-Sylo.
\end{acknowledgments}

\bibliographystyle{apsrev}
\bibliography{wboson}

\section{Supplementary material for "Loschmidt echo and the many-body orthogonality catastrophe in a qubit-coupled Luttinger liquid"}

\setcounter{equation}{0}
\renewcommand{\theequation}{S\arabic{equation}}

\setcounter{figure}{0}
\renewcommand{\thefigure}{S\arabic{figure}}

\section{Derivation of the time evolution operator}

The time evolution operator of a given mode is
\begin{gather}
U_q(t)=\exp[C_+(q,t) K_+(q)]\exp[C_0(q,t) K_0(q)]\times\nonumber\\
\times\exp[C_-(q,t)K_-(q)+i\varphi_q(t)],
\label{timeevolopsupp}
\end{gather}
where the operators
\begin{gather}
K_0(q)=\frac{b^+_qb_{q}+b_{-q}b^+_{-q}}{2},\\
 K_+(q)=b^+_qb^+_{-q}, \hspace*{8mm}  K_-(q)=b_qb_{-q}
\end{gather}
 are the generators of a
SU(1,1) Lie algebra, satisfying $[K_+(q),K_-(q)]=-2K_0(q)$, $[K_0(q),K_\pm(q)]=\pm K_\pm(q)$, and the operators
for distinct $q$'s commute with each other.

Using these and a faithful matrix representation of the generators of SU(1,1)~\cite{gilmore}, the coefficients in Eq. \eqref{timeevolopsupp} of a given mode 
 are determined from
 substituting this to the Schr\"odinger
equation of the time evolution operator, $i\dot{U}(t)=HU(t)$, or rather $i\dot{U}(t)U^*(t)=H$, where $\dot{U}(t)=\partial_tU(t)$.
Using the SU(1,1) commutation relations, we derive the identities
\begin{gather}
\exp(aK_+)K_0=(K_0-aK_+)\exp(aK_+),\\
\exp(bK_0)K_-=K_-\exp(bK_0-b),\\
\exp(aK_+)K_-=(K_-+a^2K_+-2aK_0)\exp(aK_+).
\end{gather}
Using these, we finally get 
\begin{gather}
-\dot{C}_+(q,t)+C_+(q,t)\dot{C}_0(q,t)-\nonumber\\
-\exp[-C_0(q,t)]C_+^2(q,t)\dot{C}_-(q,t)=ig(q,t),\\
i\dot{C}_-(q,t)\exp[-C_0(q,t)]=g(q,t),\\
-\dot{C}_0(q,t)+2C_+(q,t)\exp[-C_0(q,t)]\dot{C}_-(q,t)=2i\omega(q,t),\\
\dot{\varphi}_q(t)=\omega(q,t),
\end{gather}
with initial conditions $C_+(q,0)=C_0(q,0)=C_-(q,0)=\varphi_q(0)=0$.
These can be solved following Ref. \cite{truax}
as
\begin{gather}
C_+(q,t)=\frac{v^*(q,t)}{u^*(q,t)},\hspace*{4mm} C_-(q,t)=-\frac{v(q,t)}{u^*(q,t)}\\
C_0(q,t)=-2\ln(u^*(q,t)), \hspace*{2mm} \varphi_q(t)=\int\limits_0^t\omega(q,t')dt',
\end{gather}
where the time dependent Bogoliubov coefficients stem from\cite{doraquench}
\begin{gather}
i\partial_t\left[\begin{array}{c}
u_q(t)\\
v_q(t)\end{array}\right]=\left[\begin{array}{cc}
\omega(q,t) & g(q,t)\\
-g(q,t) & -\omega(q,t)
\end{array}\right]
\left[\begin{array}{c}
u_q(t)\\
v_q(t)\end{array}\right]
\end{gather}
with initial condition 
\begin{gather}
\left[\begin{array}{c}
u_q(0)\\
v_q(0)
\end{array}\right]=
\left[\begin{array}{c}
1\\
0
\end{array}\right].
\end{gather}

\section{Bogoliubov coefficients after a bang-bang protocol}

\begin{figure}[h!]
\includegraphics[width=8.5cm]{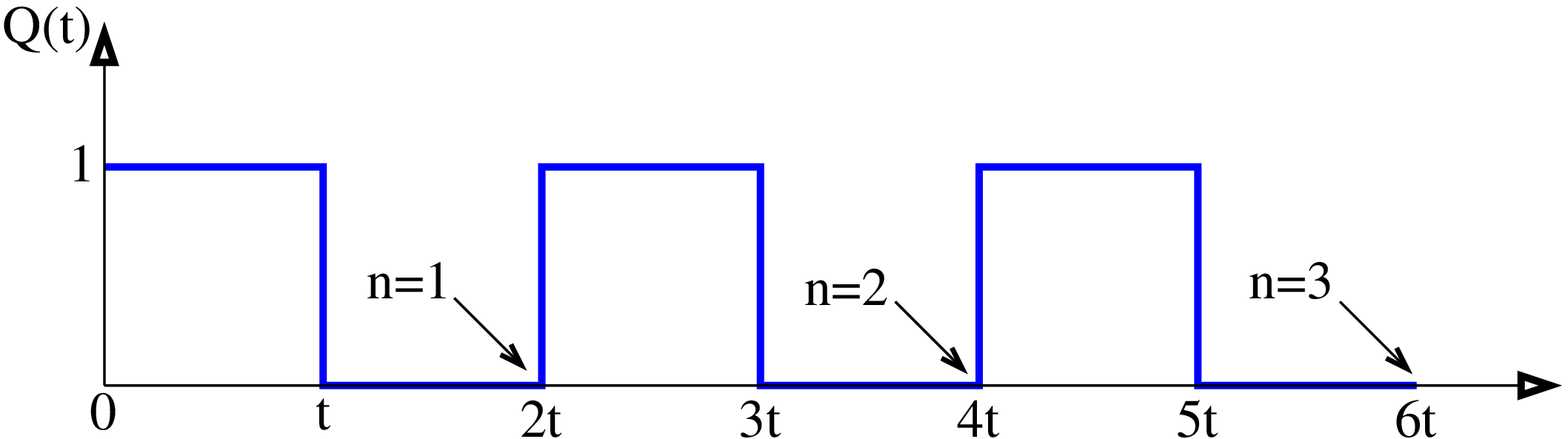}
\caption{(Color online) The bang-bang protocol is visualized up to $n=3$. The arrows indicate the time instant when the overlap is calculated
 for an $n$th order
protocol.}
\label{bangbang}
\end{figure}

Using the Bogoliubov coefficients after a SQ (i.e. $u_q(t)$ and $v_q(t)$), the multiple quench scenario, shown in Fig. \ref{bangbang} can be analyzed. 
The time dependent  Bogoliubov coefficients at time $t$ after the $n$th quench, $u^{(n)}_q(t)$ and $v^{(n)}_q(t)$ 
are determined  from
\begin{gather}
\left[\begin{array}{c}
u^{(n)}_q(t)\\
v^{(n)}_q(t)
\end{array}\right]=U^n(t)
\left[\begin{array}{c}
1\\
0
\end{array}\right],\\
U(t)=\left[\begin{array}{cc}
e^{-i\omega_i(q)t} & 0 \\
0 &  e^{i\omega_i(q)t}
\end{array}\right]
\left[\begin{array}{cc}
u_q(t) &  v^*_q(t) \\
v_q(t) &  u^*_q(t)
\end{array}\right],\\
\end{gather}
where that last interaction change occurs for an $n$th order protocol at $(2n-1)t$ as shown in Fig. \ref{bangbang},
 the time evolution afterwards only contributes with an irrelevant phase factor
to the overlap, but it is important to retain it to check the convergence of the numerics.

This leads to
\begin{gather}
\label{unt}
|u^{(n)}_q(t)|^2=1+\left|\frac{v_q(t)}{2y_q(t)}\right|^2\left[\sum_{\sigma=\pm 1}\sigma \left(s_q(t)+\sigma y_q(t)\right)^n\right]^2,
\end{gather}
where
\begin{gather}
y_q(t)=\sqrt{s^2_q(t)-1}, \textmd{ }s_q(t)=\textmd{Re}\left(u_q(t)e^{-i\omega_i(q)t}\right),\\
u_q(t)=\cos(\omega_f(q)t)-\frac i2\sin(\omega_f(q)t)\left(\frac{K_i}{K_f}+\frac{K_f}{K_i}\right)
\end{gather}
and $|v_q(t)|^2=|u_q(t)|^2-1$.
By plugging Eq. \eqref{unt} to Eq. (9) in the main text, expanding the logarithm, then performing the momentum integral 
in the $t\gg \alpha/v_f $ limit, and finally resumming of the   resulting series gives
\begin{gather}
\mathcal L_{n}(2nt)=\mathcal L_{SQ}^{n}(t)=\mathcal L_{ad}^{2n}.
\end{gather}

\section{Additional properties of the Loschmidt echo}

From Eqs. (10) and (13) in the main text, the LE is found to be symmetric with respect to the initial and final LL parameters in the steady state ($t\rightarrow\infty$)
for adiabatic and sudden quenches
as $(K_i,K_f)\longleftrightarrow (K_f,K_i)\longleftrightarrow (1/K_i,1/K_f)$.
Since these hold true in the two extreme cases,
it is plausible to assume that they remain valid for any smooth, monotonous protocol.
Throughout the calculations, we have also assumed $L>vt$, otherwise revivals show up with a period $L/2v$.

Our adiabatic and SQ overlaps parallel closely to Anderson's orthogonality catastrophe and the related x-ray edge singularity, respectively. However, while the overlap
decays exponentially with the system size in our case, the impurity problem yields a power law suppression of the overlap with the system size or time.

\end{document}